\RequirePackage{fix-cm}
\documentclass[twocolumn,epjc3]{svjour3}  

\smartqed  
\RequirePackage{graphicx}
\usepackage{color}

\usepackage{lineno}

\newcommand{\nuc}[2]{$^{#2}\rm #1$}

\newcommand{\bb}[1]{$\rm #1\nu \beta \beta$}

\newcommand{\baseT}[2]{\mbox{$#1\cdot10^{#2}$}}
\newcommand{\baseTsolo}[1]{$10^{#1}$}

\newcommand{\gline}{$\gamma$-line}
\newcommand{\glines}{$\gamma$-lines}

\newcommand{\grays}{$\gamma$-rays}

\newcommand{\tab}{Tab.~}
\newcommand{\eq}{Eq.~}
\newcommand{\fig}{Fig.~}

\journalname{Eur. Phys. J. C}
\begin{document}

\title{First limits on double beta decays in $^{\bf 232}$Th
}


\author{M. Laubenstein\thanksref{e1,addr1}
        \and
        B. Lehnert\thanksref{e2,addr2} 
        \and
       S. S. Nagorny\thanksref{e3,addr3} 
}

\thankstext{e1}{e-mail: matthias.laubenstein@lngs.infn.it}
\thankstext{e2}{e-mail: bjoernlehnert@lbl.gov}
\thankstext{e3}{e-mail: sn65@queensu.ca}


\institute{INFN - Laboratori Nazionali del Gran Sasso, 67100 Assergi (AQ), Italy \label{addr1}
           \and
           Nuclear Science Division, Lawrence Berkeley National Laboratory, Berkeley, CA 94720, U.S.A. \label{addr2}
           \and
           Queen's University, Physics Department, Kingston, ON, K7L 3N6, Canada \label{addr3}
}

\date{Received: date / Accepted: date}

\maketitle

\begin{abstract}
As one of the primordial radioactive isotopes, \nuc{Th}{232} mainly undergoes $\alpha$-decay with a half-life of \baseT{1.402}{10}~yr. 
However, it is also one of 35 double beta decay candidates in which the single $\beta$-decay is forbidden or strongly suppressed.  
181~mg of thorium contained in a gas mantle were measured in a HPGe well-detector at the Gran Sasso Underground Laboratory (LNGS) with a total exposure of 3.25~g$\times$d.

We obtain half-life limits on all double beta decay modes of \nuc{Th}{232} to excited states of \nuc{U}{232} on the order of \baseTsolo{11-15}~yr. 
For the most likely transition into the 0$^+_1$ state we find a lower half-life limit of  \mbox{\baseT{6.3}{14}~yr\ (90\%\ C.I.)}. 
These are the first constraints on double beta decay excited state transition in \nuc{Th}{232}.

\keywords{double beta decay \and excited states \and gamma spectroscopy}
\end{abstract}

\section{Introduction}
\label{intro}

Double beta decay (DBD) is a second order weak nuclear decay and subject to intense study. 
While the Standard Model process of two neutrino double beta (\bb{2}) decay is experimentally observed in 11 out of 35 possible DBD nuclides, the lepton number violating process of zero neutrino double beta (\bb{0}) decay remains elusive to date. The latter would have profound implications for particle physics and cosmology, implying the Majorana nature of the neutrino and allowing to understand the matter-antimatter asymmetry in the Universe via Leptogenesis \cite{LNV_BMV}. 

Even though the \bb{2} and \bb{0} modes require fundamentally different physics, they are connected through the same experimental techniques and share common challenges for nuclear theory. In order to interpret experimentally measured decay rates as a new lepton number violating process, nuclear matrix elements (NME) are required which are notoriously difficult to calculate. These calculations can be improved and tested by any additional experimental information of observable \bb{2} decays. 

The most likely transition for DBD is into the ground state of the daughter nucleus which is typically a \mbox{$0^+ -0^+$} transition. However, if the Q-value of the isotope is large enough, also transitions into excited state can occur. Especially useful for testing nuclear models are observed decay rates for the ground and excited states of the same nucleus. Comparing both rates, cancels many poorly constraint model parameters and allows for a more direct test of nuclear theory \cite{ESReview}. 

The end of 20$\rm ^{th}$ century and the first quarter of the 21$\rm ^{st}$ century could be considered as a ``golden age'' for direct counting experiments looking for DBD. Many experiments exploiting various detector techniques were proposed and realized within this time period.
The highest sensitivities were achieved with the ``source$=$detector'' approach, where the isotope of interest is embedded into the material of the detector. In most cases, the experimental signature is the simple sum energy of the two electrons even though some techniques aim at more advanced topology identification \cite{NEXT}.
Leading experiments reach half-life limits and sensitivities of over \baseTsolo{26}~yr \cite{KLZ0nbb,Gerda0nbb}. However, this way only certain DBD isotopes can be investigated which occur in elements suitable for a working detector technology.  

On the other hand, the ``source$\neq$detector" approach, where e.g.\ a sample containing the isotope of interest is placed on a High Purity Germanium (HPGe) detector, can be applied to searches for DBD in virtually any isotope. The ground state transitions are not accessible with this technique and the experimental signature are the de-excitation \grays\ from excited state transitions. 
Consequently, the \bb{2} and \bb{0} modes cannot be distinguished since the electrons remain in the external sample\footnote{An exception are the NEMO and SuperNEMO experiments in which a thin target foil is sandwiched between ionization chambers \cite{SuperNEMO}. }.
Such experiments have typically a smaller detection efficiency, not exceeding a few \%, and about two orders of magnitudes lower sensitivity but benefit from very unique experimental signatures of multiple \grays. The best limit with this technique was achieved for the DBD of \nuc{Se}{82} to the first excited $0^+_1$ state with \baseT{T_{1/2}>3}{22}~yr \cite{SeLimit}.
%
%
Also a combination of these two concepts is used in large scale segmented ``source$=$detector'' experiments such as GERDA, CUORE and CUPID where the decay occurs in one detector and the \grays\ are detected in another.
These searches have half-life sensitivities of about \baseTsolo{23-24}~yr \cite{GerdaES,CuoreES,CupidES}, but are again limited to certain isotopes within the detection technique.

Measuring samples on a HPGe detector in the ``source$\neq$detector" approach resulted in the first and only \bb{2} decay transitions into excited states in \nuc{Nd}{150} and \nuc{Mo}{100} with measured world average half-lives of \baseT{1.33^{+0.45}_{-0.26}}{20}~yr and \baseT{5.9^{+0.8}_{-0.6}}{20}, respectively \cite{ESAverage}.

Lower limits for excited state transitions in other DBD isotopes which were established in 
Pd isotopes \cite{searchPd},
Ce isotopes \cite{searchCe}, 
\nuc{Zr}{94} \cite{searchZr},
\nuc{Er}{162} \cite{searchYb},
\nuc{Yb}{168} \cite{searchYb},
Sm isotopes \cite{searchSm}, 
\nuc{Se}{74} \cite{searchSe74}, 
and \nuc{Hf}{174} \cite{searchHf}
within the last 5 years with half-life values in the range from \baseTsolo{17} to \baseTsolo{20}~yr.

Most of the investigated isotopes are ``classical" DBD emitters, where a nucleus $A(Z,N)$ cannot undergo single beta-decay to $A(Z\pm1,N\pm1)$ because it is energetically forbidden or heavily suppressed by an unfavorable isospin configuration. 
%
However, some of these classical DBD emitters can decay via other modes. Recent measurements with a platinum sample demonstrated a search for DBD in the unstable \nuc{Pt}{190} nuclide, which has a more favorable decay through conventional $\alpha$-decay with significantly shorter half-life. 
Systems with $\beta\beta$-processes in unstable nuclides were discussed in \cite{unstableDBD}, where lower limits on DBD of primordial \nuc{U}{235}, \nuc{U}{238}, \nuc{Th}{232} nuclides and their daughters were established. 
The authors have been utilizing long term low-background measurements with CaWO$_4$, $^{116}$CdWO$_4$ and Gd$_2$SiO$_5$ scintillating crystals for these analyses. 
The isotopes of interest were determined as internal contamination of these scintillating crystals. Despite a very low concentration of the isotopes of interest, half-life limits in the range \baseTsolo{11-12}~yr were set for the first time for the \bb{0} and \bb{2} decay modes to the ground state.

\begin{figure*}
  \centering
  \includegraphics[width=0.7\textwidth]{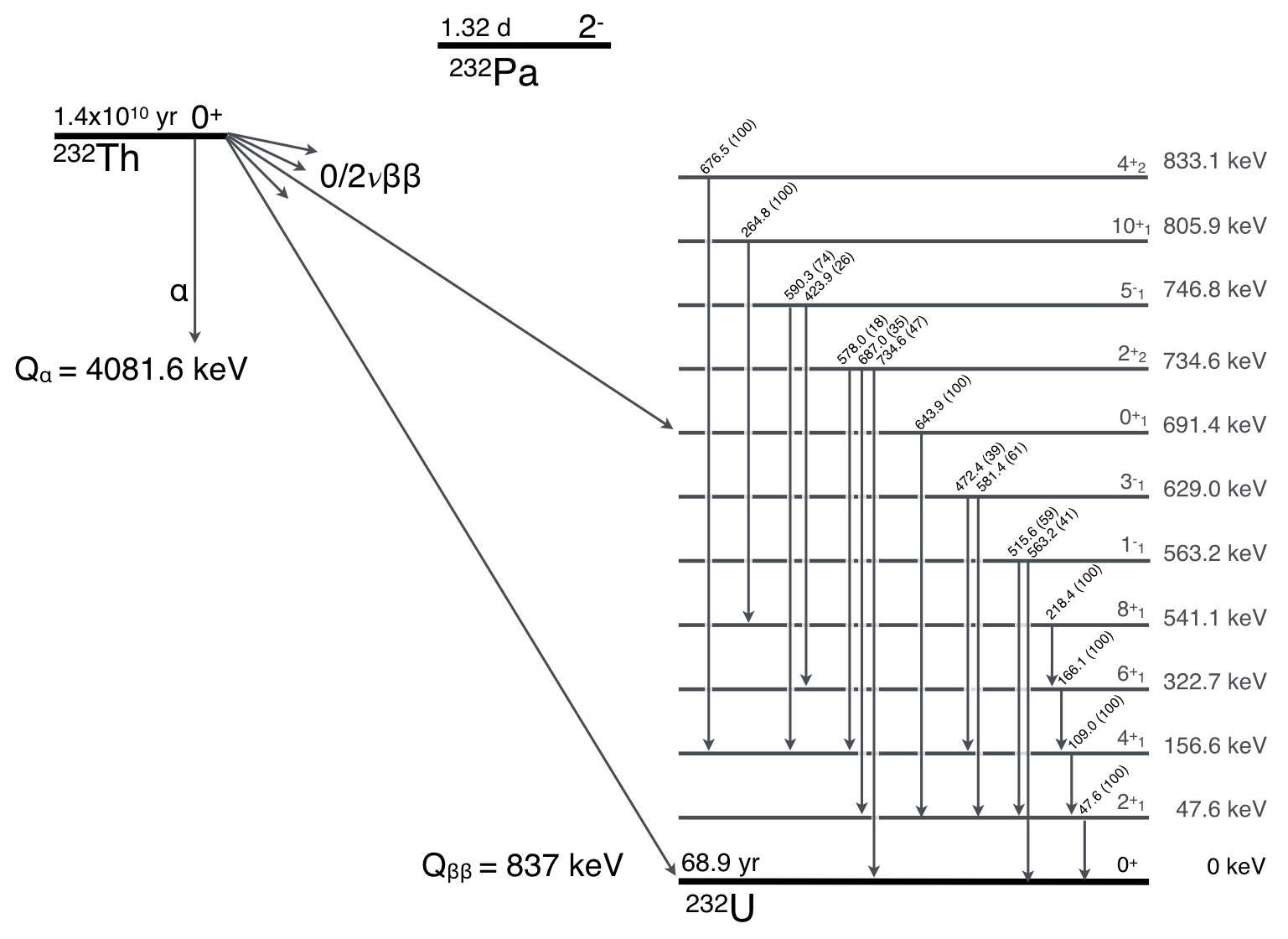}
\caption{Decay scheme of all possible $^{232}$Th double beta decay transitions. The $0^+$ transitions are highlighted. Data taken from \cite{NuclData}.}
  \label{pic:decayScheme}
\end{figure*}

In this work we investigate DBD of \nuc{Th}{232} with the ``source$\neq$detector" approach using HPGe $\gamma$-spectroscopy. 
Thorium is a mono-isotopic element and thus, the isotopic abundance of \nuc{Th}{232} is 100\% in natural thorium.
The single $\beta$-decay of \nuc{Th}{232} to \nuc{Pa}{232} is energetically forbidden but the $\alpha$-decay to \nuc{Ra}{228} is possible with \baseT{1.402}{10}~yr half-life. DBD of \nuc{Th}{232} is possible into the ground state as well as into a variety of excited states of \nuc{U}{232}. The possible decay modes are illustrated in \fig \ref{pic:decayScheme}. The most likely excited state transition is the $0^+_1$ state at 691.4~keV.
To our knowledge there were no previous attempts to search for DBD excited state transitions in \nuc{Th}{232}.

\section{Sample and Experimental Setup}

\begin{figure}
  \centering
  \includegraphics[width=0.4\textwidth]{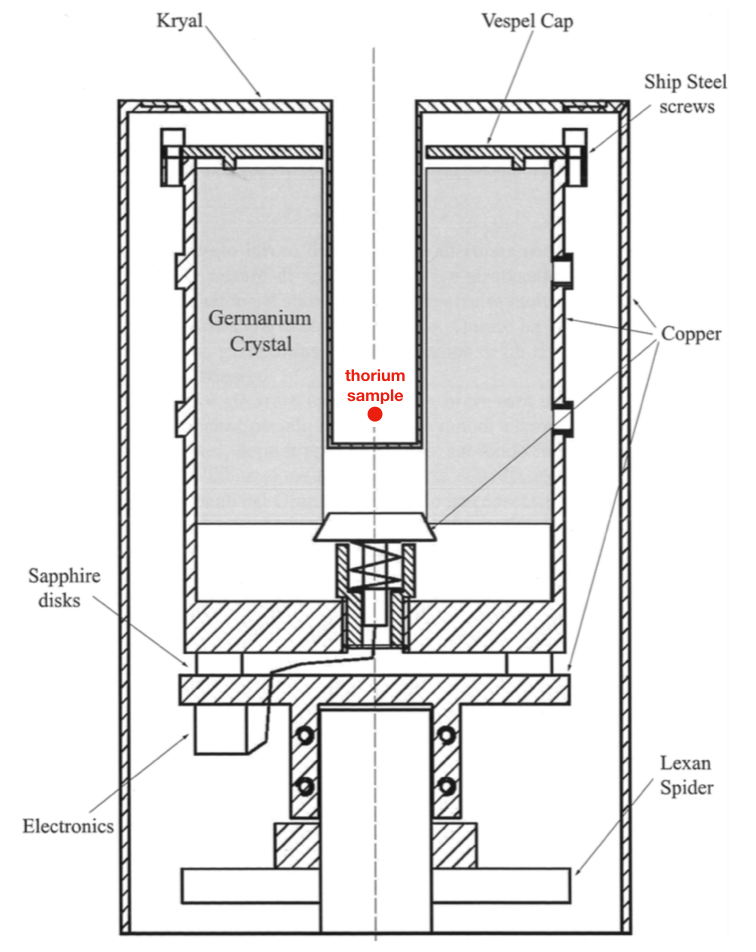}
\caption{Setup and sample configuration of the measurement. Figure adopted from \cite{detector}.}
  \label{pic:Setup}
\end{figure}

The measurement of a gas mantle sample containing 0.1811(5)~g thorium was carried out in the STELLA (SubTerranean Low Level Assay) facility in the underground laboratories of LNGS (Laboratori Nazionali del Gran Sasso) of INFN in Assergi, Italy, which provided an average shielding of $\approx3600$~m.w.e.~. Details can be found in \cite{detector,setup2,setup3,setup4}. 
The sample was placed in a 1~ml plastic vial with a cone-shaped bottom, then vacuum sealed in two plastic bags and placed into the well of an ultra low-background high purity germanium (HPGe) well-type detector. 
The HPGe detector, named GeDSG, has a 35.2\% efficiency relative to a $3\times3$ in NaI(Tl) crystal scintillator and a thin 0.75~mm aluminum window \cite{detector}. 
The detector is surrounded by a composite shield starting on the outside with 10~cm low activity lead ($< 100$~Bq/kg of \nuc{Pb}{210}), followed by another 5~cm of even lower activity lead ($< 6$~Bq/kg of \nuc{Pb}{210}) and then 5~cm of oxygen-free high conductivity (OFHC) copper, exposed only for a very short time to cosmic rays above ground. 
%
Finally, the shield and detector are enclosed in an air tight housing kept at slight overpressure and continuously flushed with boil-off from liquid nitrogen to prevent and remove radon gas from the setup. 
An illustration of the setup is shown in \fig \ref{pic:Setup}.

\begin{figure*}
  \centering
  \includegraphics[width=0.99\textwidth]{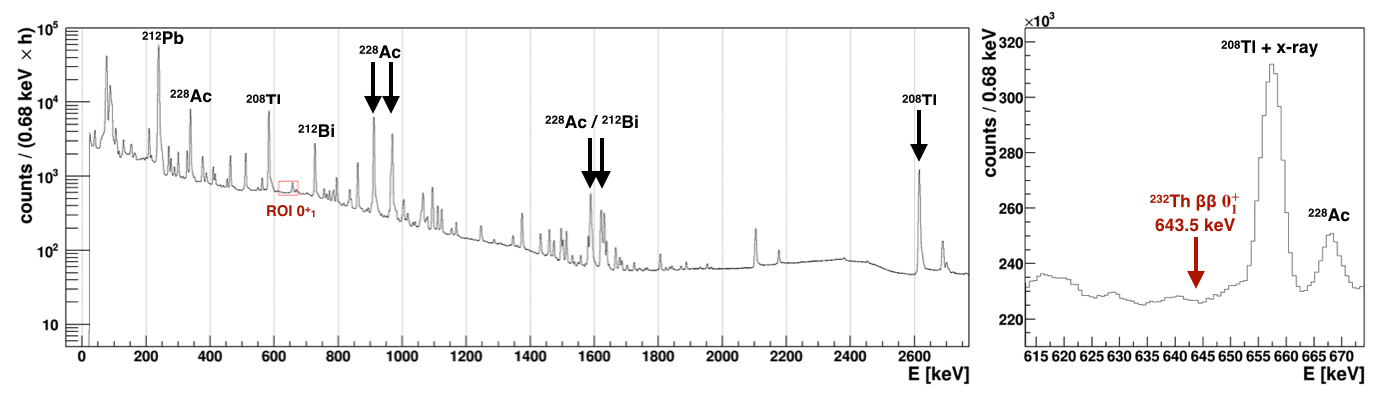}
\caption{Measured HPGe spectrum of the 0.181~g thorium sample obtained in 15.8 days. Left: full spectrum with prominent \glines\ highlighted. Right: zoom in to the region of interest for the 643.5~keV \gline\ of \nuc{Th}{232} double beta decay transition to the $0^+_1$ state. Significant spectral substructures are clearly visible in-between the main peaks due to the large number of recorded events. }
  \label{pic:WideSpec}
\end{figure*}

The energy spectrum of the thorium sample was accumulated over 378.1~h, and is presented in \fig \ref{pic:WideSpec}.
The energy resolution of the spectrometer is nominally calibrated at 2.0~keV FWHM at 1332~keV from \nuc{Co}{60}; however, the high trigger rate of the detector of about 800~Hz can deteriorate the nominal resolution performance. Thus, the abundance of \glines\ in the \nuc{Th}{232} spectrum was used for an in-situ calibration. The calibrated resolution used in the analysis is 3.7~keV FWHM at 1332~keV.
The efficiencies for the full-energy absorption peaks used for the quantitative analysis were obtained by Monte-Carlo simulation with the MaGe code based on the GEANT4 software package  \cite{bos11,moreMC}.

\section{Analysis}

The analysis is based on peak searches for de-excitation \grays\ from the various excited state decay modes. The full spectrum is shown in \fig \ref{pic:WideSpec} (left).
The high trigger rate of the detector results in \baseTsolo{4-6} counts per 0.68~keV bin which requires the search for rare events on top of a large background. The background expectation in such searches is typically taken from a background model built by Monte Carlo simulations or empirically by assuming a linear behavior around the peak. However, given the large number of events per bin, the background expectation requires per cent or even per mille precision which is not realistically achievable. \fig \ref{pic:WideSpec} (right) shows a zoom into the spectrum around the 643.5~keV \gline\ of the $0^+_1$ transition clearly indicating that a linear behavior cannot be assumed with the required precision. 
Thus, we obtain count limits of the signal peaks by excluding Gaussian peak shapes on top of the observed number of events without assuming an a-priori background. This method does not allow to discover a signal since all observed events are interpreted as background. 

The peak fits are performed in a Bayesian regime, exemplarily described for the $0^+_1$ decay mode and then applied to all possible double beta decay modes of \nuc{Th}{232}. 
The likelihood $\mathcal{L}$ is defined as the product of the Poisson probabilities over each bin $i$ for observing $n_{i}$ events while expecting $\lambda_{i}$, in which $\lambda_{i}$ is the sum of the signal $S_i$ and background $B_i$ expectation: 

\begin{eqnarray}
\label{eq:Likelihood}
\mathcal{L}(\mathbf{p}|\mathbf{n}) =
 \prod  \limits_i \frac{\lambda_{i}(\mathbf{p})^{n_{i}}}{n_{i}!} e^{-\lambda_{i}(\mathbf{p})}\ \ ,\ \  \lambda_{i}(\mathbf{p}) = S_i + B_i\ .
\end{eqnarray}

Here \textbf{n} denotes the data and \textbf{p} the set of floating parameters.

$S_i$ is taken as the integral of the Gaussian peak shape in this bin given the total signal peak counts $s$
\begin{eqnarray}
S_i&=&
 \int_{\Delta E_{i}}  \frac{s}{\sqrt{2\pi}\sigma_E} 
\cdot \exp{\left(-\frac{(E-E_0)^2}{2\sigma_E^2}\right)} dE\ , \label{eq:Si}
\end{eqnarray}

where $\Delta E_{i}$ is the bin width, $\sigma_E$ the energy resolution and $E_0$ the \gline\ energy as the mean of the Gaussian.

$B_i$, the background expectation, is implemented as a free parameter for each bin with a Gaussian prior with mean $n_i$ and width  $\sqrt{n_i}$
\begin{eqnarray}
B_i &=&
 n_i \cdot \frac{1}{\sqrt{2\pi n_i}} 
\cdot \exp{\left(-\frac{(b_i-n_i)^2}{2 n_i}\right)}\ . \label{eq:Bi}
\end{eqnarray}

This method adds an additional fit parameter for each bin but correctly distinguishes between the background expectation $b_i$ in the fit and the observed numbers of events $n_i$ on which the expectation is based. The best fit for $b_i$ will be identical to $n_i$, but the additional degrees of freedom widens the posterior distribution and results in half-life limits which are about 30\% more conservative compared to simply fixing $b_i \equiv n_i $ in the analysis.

The signal counts are connected with the half-life $T_{1/2}$ of the decay mode as
\begin{eqnarray}
\label{eq:HLtoCounts}
s =
\ln{2} \cdot  \frac{1}{T_{1/2}} \cdot \epsilon \cdot N_A \cdot T \cdot m \cdot f  \cdot \frac{1}{M}\ ,
\end{eqnarray}

where $\epsilon$ is the full energy peak detection efficiency,
$N_A$ is the Avogadro constant,
$T$ is the live-time (15.75~d), 
$m$ is the mass of the sample (0.181~g), 
and $f$ is the isotopic fraction of \nuc{Th}{232} (100\%) 
and $M$ its molar mass (232).

Each free parameter in the fit has a prior associated. The prior for the inverse half-life $(T_{1/2})^{-1}$ is flat. Priors for energy resolution, peak position and detection efficiencies are Gaussian, centred around the mean values of these parameters. The width of these Gaussians are the uncertainty of the parameter values. 
This naturally includes the systematic uncertainty into the fit result.

The uncertainty of the peak positions are set to 0.1~keV.    
The energy scale and resolution is obtained with the \nuc{Th}{232} decay chain \glines\ in the spectrum. 
A resolution of $\sigma=1.48$~keV was determined at 643.5~keV with an estimated uncertainty of 10\% which also accounts for slightly non-Gaussian peak shapes due to pile-up from the high rate operation.
The full energy peak detection efficiencies are determined with Geant4 Monte-Carlo simulations and are 14.4\% at 643.5~keV with an assumed uncertainty of 10\%. 
Systematic uncertainties on the measured sample mass and the isotopic fraction in the sample are negligible with respect to the uncertainty of the detection efficiency.

The posterior probability distribution is calculated from the likelihood and prior probabilities with the Bayesian Analysis Toolkit (BAT) \cite{Caldwell:2009kh} and marginalized for $(T_{1/2})^{-1}$. 
The best fit is always zero signal counts in this method since all observed events are consistent with the background by design. The 90\% quantile of the marginalized posterior distribution of $(T_{1/2})^{-1}$ is used to set the 90\% credibility limits including systematic uncertainties. For the $0^+_1$ transition, 3145 counts are excluded in the 643.5~keV peak on top of a background of \baseT{3.5}{5}~cts/keV. The lower half-life limit is
\begin{eqnarray}
\label{eq:ExampleHL}
T_{1/2}  > 6.7\times 10^{14}\ {\rm yr}\ (90\%\ {\rm CI}).  
\end{eqnarray}

\begin{figure}
  \centering
  \includegraphics[width=0.4\textwidth]{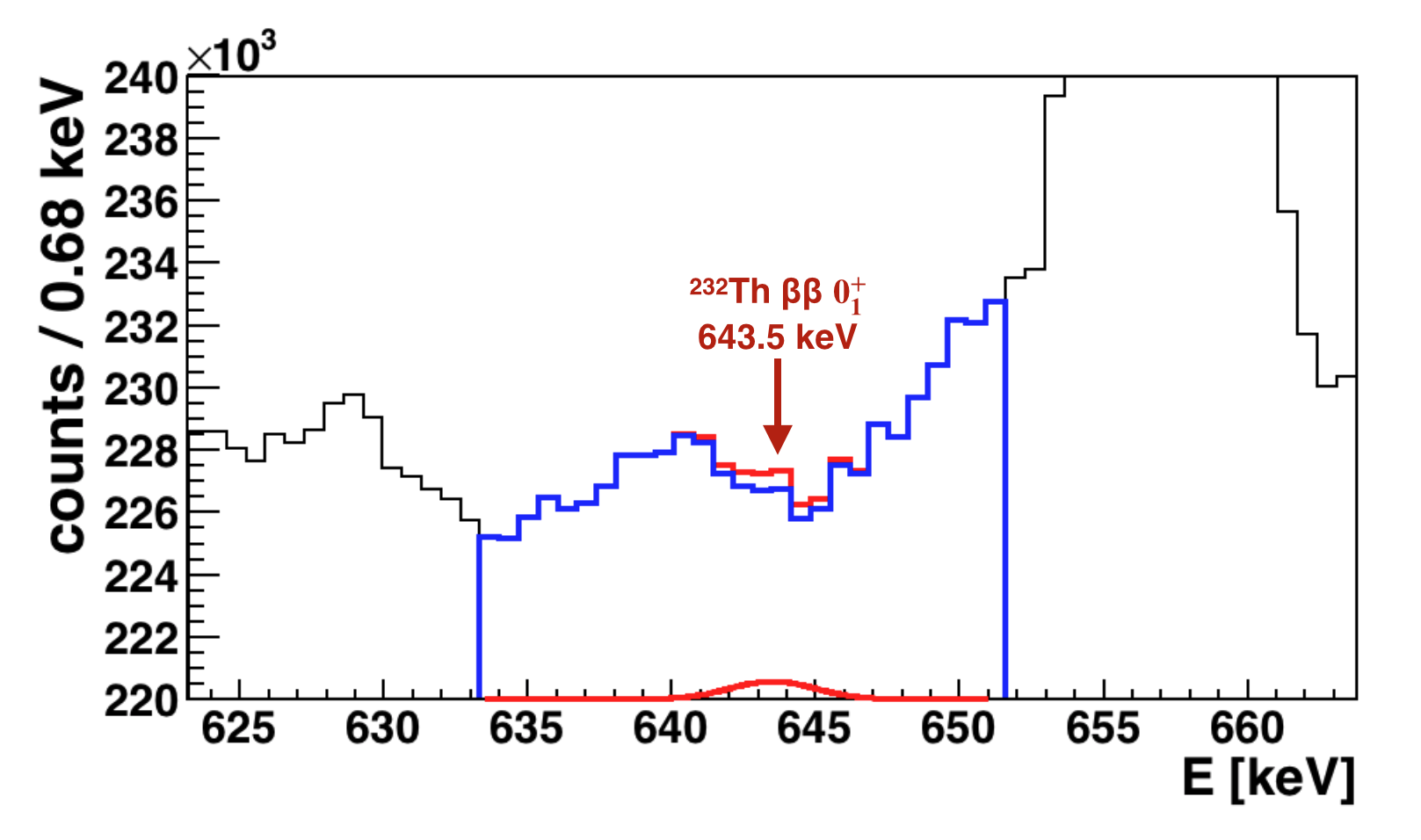}
\caption{Region of interest and fit for the $0^+_1$ transition. The data is shown in black. The mean background expectation in each bin identical to the data by construction is shown in blue. The signal peak excluded by 90\% probability is shown in red on of the data as well as independently at the bottom.}
  \label{pic:ROIFit}
\end{figure}

The fit is shown in \fig \ref{pic:ROIFit} illustrating the fit function in red with the signal peak set to the strength excluded with 90\% credibility. The fit function for the best fit, i.e.\ without signal strength and background equivalent to the observed number of events, is shown in blue.

The other decay modes are treated similarly and results are shown in \tab \ref{tab:HLimits}. In case multiple \glines\ are considered, a combined fit is performed by extending the likelihood in \eq \ref{eq:Likelihood} over multiple regions of interest with common $(T_{1/2})^{-1}$ parameter. The 47.6~keV \gline\ is part of all decay modes but due to its low energy it has only a small detection efficiency and higher background level in the setup. It is only considered for the first excited state ($2^+_1$) where it is the only \gline\  and where the obtained half-life limit is about 3 orders of magnitude smaller than for the other modes. The 578.0~keV \gline\ of the $5^-_1$ state is omitted due to its low branching ratio. The complete list of considered \glines\ is also presented in \tab \ref{tab:HLimits} for each decay mode.


\begin{table}[htbp]
\begin{center}
\begin{tabular}{lll}
\hline
decay level        & T$_{1/2}$ (90\% CL)   & \glines\ energies \\
 \  [keV] ($J^\pi $)    &  [yr]     & [keV]  \\
\hline
 47.6 $(2^+_1)$   &  \baseT{>4.8}{11} &  47.6                       \\
156.6 $(4^+_1)$   &  \baseT{>9.1}{13} &  109.0                      \\
322.7 $(6^+_1)$   &  \baseT{>3.2}{14} &  109.0, 166.1               \\
541.1 $(8^+_1)$   &  \baseT{>1.8}{14} &  109.0, 166.1, 218.4        \\
563.2 $(1^+_1)$   &  \baseT{>4.4}{14} &  515.6, 563.2               \\
629.0 $(3^+_1)$   &  \baseT{>3.6}{14} &  109.0, 472.4, 581.4        \\
691.4 $(0^+_1)$   &  \baseT{>6.7}{14} &  643.5                      \\
734.6 $(2^+_2)$   &  \baseT{>3.8}{14} &  686.8, 734.6               \\
746.8 $(5^-_1)$   &  \baseT{>1.7}{14} &  109.0, 166.0, 424.3, 590.4 \\
805.9 $(10^+_1)$  &  \baseT{>4.1}{14} &  109.0, 166.0, 218.4, 264.8 \\
833.1 $(4^+_2)$   &  \baseT{>5.9}{14} &  109.0, 676.5               \\
\hline
\end{tabular}
\medskip
\caption{\label{tab:HLimits} Lower half-life limits on \nuc{Th}{232} double beta decay modes set in this work. The last column shows the \glines\ used in the combine fit.
}
\end{center}
\end{table}

\section{Conclusions}

We performed a first search for double beta decays of \nuc{Th}{232} into all possible excited states of \nuc{U}{232} using a thorium containing gas mantle sample and a HPGe well detector. 
The established limits are valid for both the \bb{2} and \bb{0} mode. 
The large intrinsic background of the experiment did not allow to model the background prediction with sufficient precision. Thus the analysis was performed without background model and limits on \nuc{Th}{232} DBDs were set under the assumption that all observed events are background, i.e.\ a discovery with this method is not possible. 


Future improvements of this measurement should aim at reducing the intrinsic background in the search. The vast majority of background originates from the decay daughters of the sample nuclide itself which build up over time. This could be reduced by using a Th sample right after the separation/purification process, for example by anion-exchange resin, so that existing daughter isotopes are chemically removed. 

The Th-daughter nuclides will then again accumulate with time. Thus the mass of the sample must be well chosen based on the acceptable count rate for the used detector setup. 
Moreover, one could consider a campaign of several subsequent runs of the same measurement, interrupted by intermittent re-purification/separation of \nuc{Th}{232} from its daughter nuclides accumulated within the previous measurement periods.

For investigating the DBD ground state transition, only the ``source$=$detector" approach can be used. In this situation it is difficult to realize campaigns with intermittent purification since the extraction of Th from any detector material is a more complicate and time-consuming process.
However, using a detector technology with high characteristic time-response, such as fast scintillators, would allow to deploy a larger amount of thorium. 
For characteristic scintillating times in the range 1-50~$\mu$s, one could load a scintillator with a few hundred kBq activity of \nuc{Th}{232}.
Taking into account that activities of its daughter will rise with time and their beta components create a large continuous background, the initial \nuc{Th}{232} amount should be reduced to a few kBq.
A potential type of detector are scintillating crystals grown by the Bridgman technique in a closed quartz ampule. This allows the whole growing setup to be safely contained and avoid contaminations from the radioactive Th-chain isotopes during production. Additional precaution must be taken during the machining and handling of such Th-loaded crystals. 
In order to avoid a possible interference between an activator in the scintillator and the \nuc{Th}{232} load, the scintillator should be self-activated.
Potential candidates are halide crystals, such as NaI (undoped), CsI, or a new type of Cs$_2$HfCl$_6$ crystals. 
%

Here, we investigated only DBDs of \nuc{Th}{232} but in principle also other radioactive DBD isotopes in the \nuc{Th}{232} decay chain can be search for, as suggested in \cite{unstableDBD}. However, given their low abundance as \nuc{Th}{232} daughters and much shorter half-lives than \nuc{Th}{232}, the resulting half-life limits would be significantly lower.

There are no existing theoretical estimations of half-lives for excited state transitions of \nuc{Th}{232}. 
Such calculations require significant effort and are limited to certain nuclear models e.g.\ QRPA, since the daughter nuclide \nuc{U}{232} is heavily deformed and has strong rotation bands. 
From the $Q_{\beta\beta}^5$ dependence of \bb{2} half-lives, one can crudely extrapolate from observed \bb{2} decays in other isotopes that the here established sensitivities are about 10 orders of magnitude lower than needed. 
The half-life of the first excited $0^+_1$ state is expected to be about 6000 times longer for the $0^+$ ground state.

Nevertheless, steady progress in improving half-life limits for DBD isotopes, other than the ones used in large scale experiments (i.e.\ \nuc{Ge}{76}, \nuc{Mo}{100}, \nuc{Se}{82}, \nuc{Te}{130}, \nuc{Xe}{136}), will eventually help to tune theoretical models to better describe the nuclear physics behind double beta decay.


\begin{acknowledgements}
We would like to thank Dr.~Fedor \v{S}imkovic for interesting and useful discussion about theoretical half-life estimates.

\end{acknowledgements}

\bibliographystyle{spphys}       


\end{document}